\documentclass{vldb}
  \usepackage{times}
\usepackage{graphicx}
\usepackage{balance}  
\usepackage{url}
\usepackage{color}
\makeatletter
\newif\if@restonecol
\makeatother

\newcommand{\mypar}[1]{\smallskip\noindent\textbf{#1}}

\usepackage[ruled,vlined]{algorithm2e}

 \newcommand{\ncaption}[1]{\caption{\small   #1 } }

\usepackage{subfigure}

\begin{document}

\newcommand{\carlos}[1]{\textcolor{red}{Carlos: #1}}
\newcommand{\aapo}[1]{\textcolor{cyan}{(Aapo: #1)}}

\newcommand{\denselist}{
  \itemsep -5pt\topsep-8pt\partopsep-8pt
}
\newcommand{\quitedenselist}{
  \itemsep -1pt\topsep-8pt\partopsep-8pt
}

\title{GraphChi-DB: Simple Design for a Scalable Graph Database System -- on Just a PC}



%
%
%
%

\numberofauthors{2} 

\author{
%
%
\alignauthor
Aapo Kyrola\\
       \affaddr{Computer Science Department}\\
       \affaddr{Carnegie Mellon University}\\
       \affaddr{Pittsburgh, PA, USA}\\
       \email{akyrola@cs.cmu.edu}
\alignauthor
Carlos Guestrin\\
       \affaddr{Computer Science Department}\\
       \affaddr{University of Washington}\\
       \affaddr{Seattle, WA, USA}\\
       \email{guestrin@cs.washington.edu}
}
 \date{March 1st 2014}


\maketitle

\begin{abstract}
We propose a new data structure, Parallel Adjacency Lists (PAL), for efficiently managing graphs with billions of edges on disk. The PAL structure is based on the graph storage model of GraphChi \cite{graphchi:osdi2012}, but we extend it to enable online database features such as queries and fast insertions. In addition, we extend the model with edge and vertex attributes. Compared to previous data structures, PAL can store graphs more compactly while allowing fast access to both the incoming and the outgoing edges of a vertex, without duplicating data.   Based on PAL, we design a graph database management system, GraphChi-DB, which can also execute powerful analytical graph computation.

We evaluate our design experimentally and demonstrate that GraphChi-DB achieves state-of-the-art performance on graphs that are much larger than the available memory.  GraphChi-DB enables anyone with just a laptop or a PC to work with extremely large graphs.
 
\end{abstract}

\section{Introduction}

In the recent years, researchers in the academia and industry have proposed various specialized solutions for handling extremely large graphs, such as social networks, in scale. Systems such as GraphLab \cite{powergraph} and Pregel \cite{pregel} can execute \emph{computation} on graphs with billions of edges on a cluster by partitioning the graph so that each machine can store a part in memory.  On the other hand, the database community has been working on specialized \emph{graph databases} such as Neo4j \cite{neo4j}, Titan \cite{titan} and DEX \cite{dex} that are designed for fast graph traversals, arguing that relational databases are not suited for the task -- argument that is not shared by all researchers\footnote{http://istc-bigdata.org/index.php/benchmarking-graph-databases/}.  Systems offering both database functionality and large-scale analytical computation for graphs include the distributed systems Kineograph \cite{kineograph}, Trinity \cite{shao2013trinity} and Grappa \cite{grappa}. Allowing analytical computation, or \emph{global queries},  to be executed directly inside a database reduces the complexity of the analytics workflow, as data scientists do not need to concern themselves with error-prone transfers of data from one system to another.

Lately, several researchers (\cite{graphchi:osdi2012,turbograph,xstream}) have demonstrated that distributed computation is not necessary for large-scale graph computation, but by using disk-based algorithms, even a single consumer PC can handle graphs with billions of edges. In this work, we build on the GraphChi system, which we proposed in \cite{graphchi:osdi2012}, to create a powerful, scalable graph database ``GraphChi-DB''. The storage engine of GraphChi-DB is based on a novel Partitioned Adjacency List (PAL) data structure which is derived from the storage model used by GraphChi  \cite{graphchi:osdi2012}. Our contributions enable fast queries over the graph and extends the model by allowing attaching arbitrary attributes for both edges and vertices of the graph. We adapt the Log-Structured Merge Tree (LSM-tree) \cite{LSM} to obtain very high insert performance. GraphChi-DB retains the computational capabilities of GraphChi. Because it is designed to handle data that is much larger than the available memory, GraphChi-DB enables working efficiently with much larger graphs on just a PC or laptop than previously possible. 

 The PAL data structure is very simple and easy to implement. There are several benefits to this simplicity: (1) the database storage occupies much less space on disk compared to other graph database management systems, allowing very large graphs to be handled on just a commodity Solid-State Disk (SSD);  (2) the data structure is based on immutable flat arrays, simplifying greatly the system design because complex mechanisms are not needed for protecting the integrity of the data structure in case of failures; (3) all indices can typically fit in memory, even on just a laptop with limited amount of memory, reducing random disk accesses and saving space compared to relational database storage. In this work, we carefully analyze the trade-offs of enabling fast writes at the expense of fast queries, and demonstrate that our design reaches a good compromise by evaluating it against state-of-the-art graph databases.  Importantly, our design is flexible and allows user to choose a model that best suits a particular workload: for example, GraphChi-DB can be used as an OLAP data warehouse for analytics or as an online graph database processing hundreds of thousands of writes per second. 

The outline of this paper is as follows: After presenting some preliminaries in Section 2,  we discuss previous approaches for storing graph data on disk in Section 3. Our main contribution, the PAL-structure is presented in Sec. 4 with analysis.  In Sec. 5, we modify the model to allow very high insert throughput.  In Section 6, we present the  model for analytical computation, based on \cite{graphchi:osdi2012}, but adapted to our modified data structure.  Finally, in Sec. 7, we discuss some of the implementation details of GraphChi-DB and provide comprehensive evaluation of the system in Sec. 8.

In short, our contributions are:
\begin{itemize}
\denselist

\item We propose a novel graph storage format, Partitioned Adjacency Lists (PAL), which addresses many of the shortcomings of previous graph storage models. In addition to allowing fast access to both in- and out-edges of vertices, the model allows efficient access to the edge and vertex attributes, without additional indices, and is also an efficient data structure for disk-based analytical graph computation.

\item We adapt the Log-Structured Merge-tree \cite{LSM} for our purposes, allowing insertion of hundreds of thousands of edges per second on just a laptop, without a special batch mode.

\item Based on PAL, we present design for a highly scalable single-node graph database, GraphChi-DB, which can also execute graph computation ``in-place". Adjusting the parameters of the PAL structure, GraphChi-DB can be tuned for various types of workloads.  We show that GraphChi-DB performs much better than existing graph databases on data that is many times larger than the available memory.  We release the software in open source: \url{https://github.com/GraphChi} (pending).

\end{itemize}

%

\section{Preliminaries}

In this work, we use the directed graph model:  Structure of a graph is defined as $G = (V, E)$ where $V$ is the set of {\bf vertices} (nodes) with integer ids. {\bf Edges} (links) $E$ is a relation $V \times V$. Thus, edge is a directed tuple $(source, destination)$ where $source, destination \in V$. For example in the Twitter's follow-graph, an edge $(a, b)$ means that $a$ ``follows'' $b$.  We call edge $e = (u, v)$ an {\bf out-edge} of vertex $u$ and  an {\bf in-edge} of vertex $v$. Similarly, $v$ is then $u$'s {\bf out-neighbor}, and $u$ is the {\bf in-neighbor} of $v$.  (In/out) {\bf degree} of a vertex is the number of incident (in/out) edges.
With each vertex and edge we can store arbitrary attributes.   We generally assume the graph to be \emph{sparse}, i.e that $|E| \ll| V|^2$.

For our analysis, we use the I/O model proposed by Aggarwal and Vitter \cite{aggarwal1988input}. In this model, we count the number of block transfers between disk and memory. The complexity is parameterized by block size $B$.  


\section{Graph Storage on Disk}

In this section we discuss briefly the established approaches to store graphs on disk and their respective advantages and disadvantages. 

\subsection{Adjacency Lists}

In the  out-directional Adjacency Lists model, all out-neighbors of a vertex are stored in a list keyed by the vertex ID: $(v, [u_1 \mapsto x_1,  u_2 \mapsto x_2, u_3\mapsto x_3, ...]) \, | \, \forall u_j \, (v, u_j) \in E$.  Values $\{ x_j \}$ refer to the edge attributes. Instead of the values themselves, we can also store a foreign key (or pointer) to a separate table that contains the data. In the in-directional model, in-neighbors are stored instead. Note that the neighbors can also be stored as a bitmap \cite{dex}, instead of a list of vertices.  Adjacency lists can be materialized on disk in various ways, the choice depending on the expected workload. 

{\bf Advantages:} Adjacency lists allow fast access to the in- \emph{or} out-neighbors of a vertex. If access to both in- and out-neighbors is required, two adjacency lists must be stored for each vertex (otherwise, finding in-neighbors from an out-adjacency list requires a full scan).  Adjacency Lists  achieve good locality of access to neighbors of a vertex, assuming each neighbor list is stored sequentially on disk (as opposed to storing them as a linked list). Adjacency list requires index of only size $O(V)$, where $index(u)$ contains pointer to the beginning of $u$'s adjacency list. Large neighbor lists can often be compressed. 

{\bf Disadvantages:} If access to both in- and out-neighbors is required, the storage requirements are doubled. Also, if both in- and out-adjacency lists are used then each edge insertion and deletion requires at least two disk accesses (finding a specific neighbor from the adjacency list may require additional disk accesses). If we store the attributes with each edge, then change of a value needs to be done in both vertices' lists (if we a store pointer/foreign key to the edge data, then only one access is needed). When executing graph computation that requires full pass over the graph and modifies the edge values, we must choose to process sequentially either in- or out-edges of each vertex, but not both \cite{graphchi:osdi2012}.

Updating the adjacency list can be costly: a common technique (also used by TurboGraph \cite{turbograph}) is to store neighbor lists in pages that have some empty space, allowing new neighbors to be added without adding new pages. However, when the extra space runs out, new pages have to be allocated and the locality of access is reduced.

\subsection{Edge List}

In the Edge List model, each edge tuple $(src, dst)$ is stored as an individual entry. For example, in a relational database each edge would be a row with columns for source and destination vertex ID. To enable fast access to all edges of a vertex (in- or out-edges or both), as required by typical graph queries, we can either (a) create an index over $src$ or $dst$; or (b) store edges in a doubly linked list (this solution is implemented by Neo4j \cite{neo4j}): each edge entry stores pointer to the previous and the next edge for both the $src$ and $dst$ vertices. An index for the first edge of each vertex is also required.

{\bf Advantages:} Each edge can be stored only once, so changing value of an edge requires only a single access. If the graph fits in memory, traversing edges of a vertex is fast pointer jumping with $O(1)$ access per hop. In the linked list model, traversing two-hop neighbors (``friends-of-friends'') can be especially fast as each edge provides direct access to the edges of the neighbor.  

{\bf Disadvantages:} Indices created over the edges often take more space than the edges themselves: for example, using MyISAM table in MySQL \cite{mysql}, storing the edge tuples takes just 9 bytes per edge but the B-tree \cite{comer1979ubiquitous} index over $src$ or $dst$ IDs takes 11 bytes / edge (on a graph with 68M edges). Updating the indices when edges are inserted or removed can be costly. Using double linked lists avoids the problem with large indices but the overhead of storing four pointers for each edge is considerable. For example, Neo4j \cite{neo4j} uses 33 bytes / edge \cite{robinson2013graph}. Also, inserting an edge requires updating pointers of the previous edges of both of the endpoint vertices. Traversing a linked list is inherently sequential.


\subsection{General Challenges of Large Graphs}

\label{sec:powerlaw}

A particular challenge to graph computation has been dealing with so called natural graphs \cite{powergraph}, for example the Web and social networks such as Twitter. The in-degree distribution of those graphs follows the power law (Zipf distribution): in layman terms, some ``hot'' nodes have extremely large number of in-neighbors but most nodes have only few. For example, in the Twitter graph, the most popular celebrities have tens of millions of followers while the median user has only some dozens. This kind of structure makes finding good partitions (cuts) of the graph extremely challenging, making efficient distributed computation hard \cite{graphchi:osdi2012, powergraph}. Similarly, we cannot generally partition graphs on disk to localize access. Highly random access patterns make efficient buffer management and caching difficult to implement. 


\newpage

%
%
%
%
%

\section{Partitioned Adjacency Lists}

 
 This section describes the Partitioned Adjacency Lists (PAL) data structure that is the basis of GraphChi-DB. The primary objective of the design is to reduce the number of random accesses while minimizing the storage space required.
 PAL has the following features: (1) it requires each edge to be stored only once, while (2) providing 
 efficient access to both in- and out-edges of a vertex; (3) edge attributes are stored in column-oriented
 storage symmetrically to the adjacency information so that fast access to the edge data does not require a foreign key index.
  Furthermore, PAL allows processing the whole graph efficiently using  (4)  the Parallel Sliding Windows (PSW) algorithm proposed in \cite{graphchi:osdi2012}.  
 In the next section, we describe how we can enable fast insertions to the graph.

 \mypar{Data model: } The PAL data model is a directed graph $G = (V, E)$ with two types of objects: \emph{vertices (V)} and \emph{edges (E)}. Vertices are identified by integer IDs and an edge is an ordered tuple $(u, v, \tau) \, |  \, u, v \ \in V$,  $\tau$ is the edge type. The set of edges encode the structure, or \emph{connectivity} of the graph. Both edges and vertices can have attributes, organized in typed \emph{columns} (edge type could also be stored as an attribute, but because graph queries usually select by edge type, it is more efficient to include the type in the edge). 
 
 For example, for a graph representing a social network, vertices could have associated columns for user's country code (char[2]), account creation date (date) and a score for user's influence (float) in the network (computed for example with the Pagerank \cite{pagerank} algorithm). An edge of type ``follows'' would mean that an user has subscribed to the messages of another user: if Jack follows Jill, there is edge from Jack's vertex to Jill's vertex (but possibly not vice versa). Edges could have columns for the edge creation time (timestamp) and a weight-column (float) that stores the strength of the relationship.  
  
\subsection{Edge Partitions}

 \begin{figure}[t]
\centering
   \includegraphics[width=0.4\textwidth]{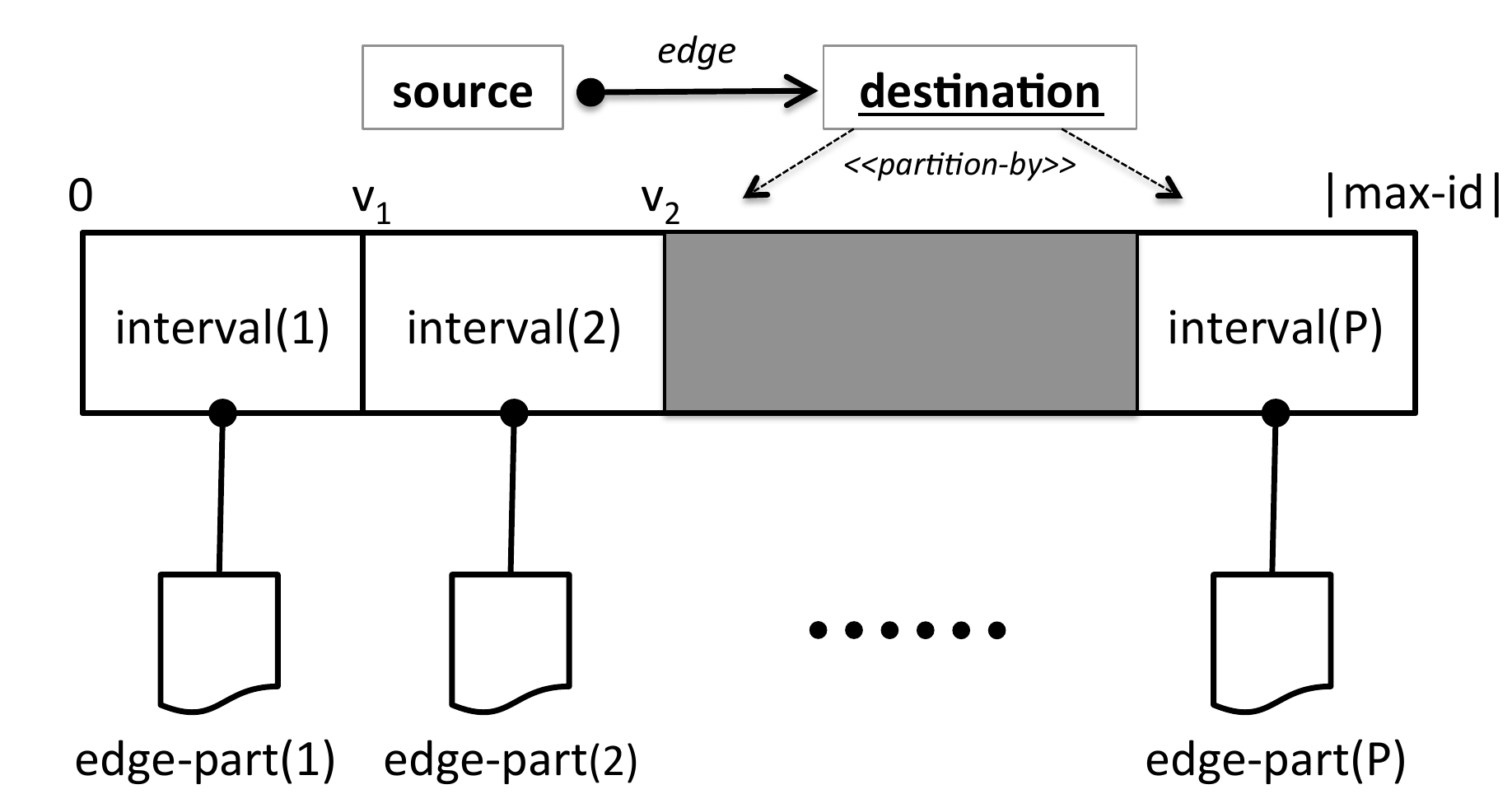}

 \ncaption{The range of vertex IDs is divided into $P$ intervals. Each interval is 
 associated with an edge partition,  which stores all the edges that have destination ID in the interval. In the partitions, edges are sorted by the source ID. }
 \label{fig:intervals}
 
 \end{figure}

  \begin{figure}[h]
\centering
   \includegraphics[width=0.4\textwidth]{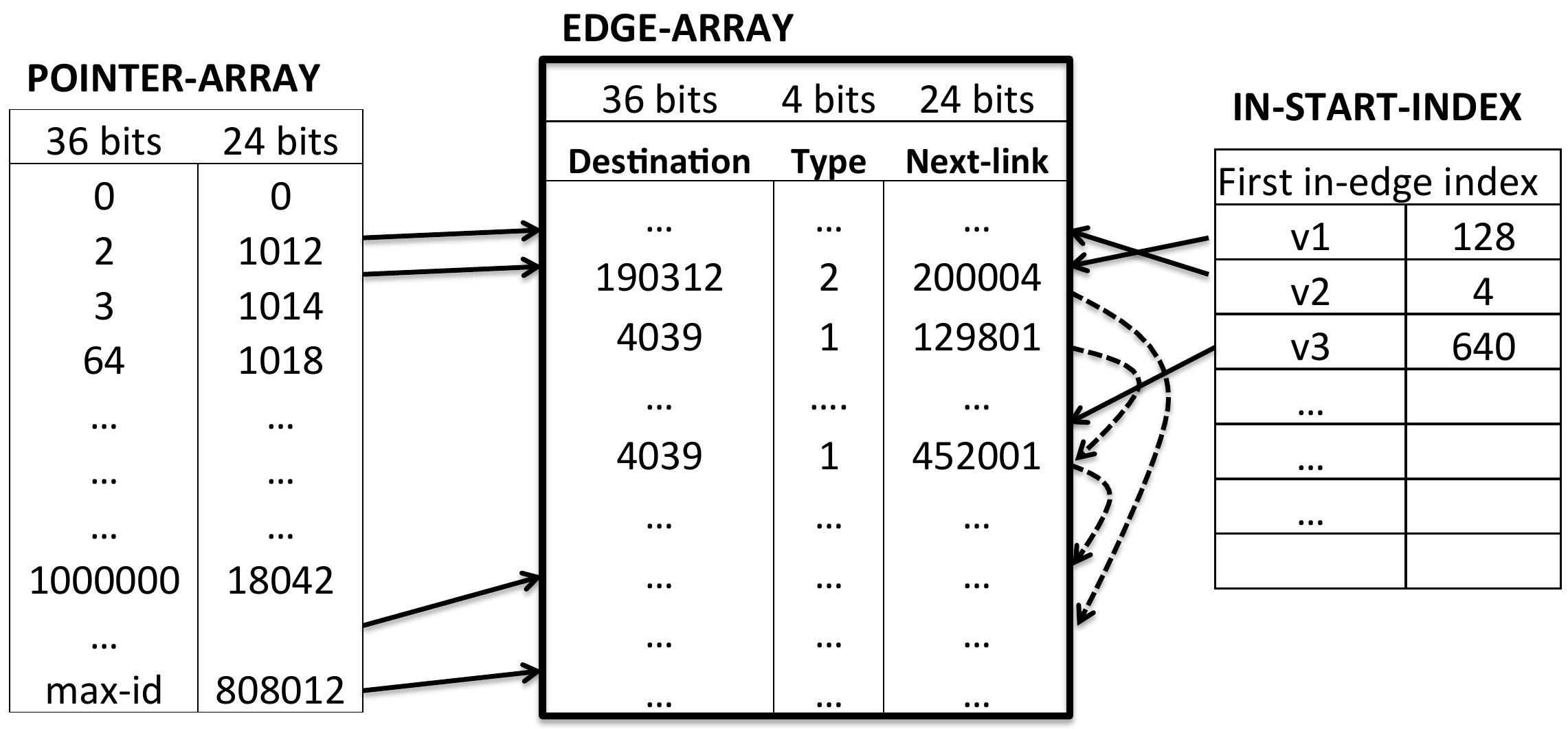}

 \ncaption{ File structure of an edge partition. }
 \label{fig:partitionstructure}
 
 \end{figure}
 
 We first describe the ``partitioning'' in Partitioned Adjacency Lists, which is exactly same as proposed in \cite{graphchi:osdi2012}. Let $[0, max-id]$ be the range of allowed vertex IDs and let $P$ be the number of partitions. We split the ID range to $P$ continuous intervals as shown in Figure \ref{fig:intervals}. With each {\bf interval(i)} we associate an {\bf edge-partition(i)}\footnote{We do not use the term ``shard'' as used by \cite{graphchi:osdi2012} to avoid confusion with the other uses of the term.} which stores all edges $(source, destination)$ such that  $ destination \in \mbox{ interval(i)}$. Importantly, the edges are stored in sorted order by the $source$ ID. Note, that the model does not require intervals to be of equal length, but intervals should be chosen so that any one edge-partition fits into memory. However, unlike in the batch computation setting, in the online setting the distribution of edges is not known beforehand. To balance the edge distribution, GraphChi-DB uses a reversible hash function to map the vertex IDs so that edges are, in expectation, distributed evenly into partitions (details in Sec. \ref{sec:vertexid}).  
 
\mypar{Remark: } The choice of the edge $destination$ ID  as the partition key, instead of the $source$ ID, is arbitrary and can be reversed without any changes to the model.

\mypar{Constraints: }Our model requires enough memory to hold any single vertex's edges in memory (see \cite{graphchi:osdi2012}). In the analysis, we assume that the number of in-edges of a vertex does not exceed $|E|/P$.  Both of these constraints could be addressed with some care, but in practice, we have not encountered graphs where these constraints would be a limitation, even when using just a few gigabytes of memory.
 
 \subsubsection{File Structure (Graph Connectivity)}
 \label{sec:structure}
 
The file structure of an edge partition for storing the graph's connectivity is based on the Compressed Sparse Row (CSR) format, which is commonly used to store sparse graphs.  Referring to the Figure \ref{fig:partitionstructure}, the main file (in the center) is a flat {\bf edge-array} with one entry for each edge. The edge encoding is not specified by PAL, but for concreteness we present the format used by GraphChi-DB. Each entry stores the {\it destination} vertex ID (36 bits), a 4-bit edge type and a 24-bit offset to the next edge with same destination ID (see below for discussion). The {\bf pointer-array} file stores the {\tt edge-array} position of the first out-edge of each vertex (in ascending order).  Sparse format is used, so only those vertices that have any out-edges in the partition have an entry.   The {\bf in-start-index} file stores the {\tt edge-array} position of the first in-edge of each vertex included in the partition, if any.  More bits can be used if the range of IDs exceeds $2^{36}$.

It is important to note that the edge partition is an \emph{immutable} data structure. The only modification we allow is changing the type of an edge, as it does not change the order of edges or the size of the file. Instead, new edges are merged in ``bulk'' to a partition by creating a new file. Insertions are described in Section \ref{sec:inserts}.

 

%
%

\subsection{Retrieval of Edges}

Graph queries are built on primitive queries that return in- or out-edges of a given query vertex. When an edge is found from the database,  the result set contains the edge tuple $(u, v, \tau)$ and the position of the edge in the edge partition. The position can be then used to access edge attributes, as explained in the next section. 
 In this section we describe how these queries are executed in the PAL model.   

\subsubsection{Out-edges}

\label{sec:outedges}

{\it  Out-edge query:  Given $u \in V$, and an edge $type$, return all
 $  (src, dst, \tau) \in E \,  | \, src = u, \, \tau=type$. For example, in a social network, for a query vertex $u=$ `Jack', find all the links \emph{to} users that Jack ``follows''. }

\smallskip

Recall that the edges are ordered by their $source$ ID in the edge partition. A vertex can have out-edges in all of the $P$ partitions.  To find the out-edges of a vertex $v$ from an edge-partition(i), we can first search the {\tt pointer-array} using binary search to find the offset $a$ to the first out-edge (if any) of the vertex, and also the offset $b$ of the first edge of the following vertex. Then, we need to read consequent values {\tt edge-array[a..(b-1)]}, which requires only one random seek to disk. Unfortunately, the cost of the binary search of the {\tt pointer-array} can be high if the file is not cached in memory. 

We propose two options to solve this problem. First, we notice that the vertex IDs and file offsets in the {\tt pointer-array} are increasing integer sequences. We can then encode the differences between subsequent values (a technique common in index compression), which are small on average. 
In our implementation, we used the Elias-Gamma coding \cite{elias1975universal}, which typically compresses the {\tt pointer-array} to only a fraction of the original size, allowing us to permanently pin the index to memory and so avoid disk access completely. 

The second approach is to compute a \emph{sparse index} of the {\tt pointer-array} that we store in memory. Then, instead of doing binary search on the full index file, we first consult the sparse index in memory to narrow the range of the search on disk. 

\mypar{ Cost analysis: } The number of random accesses is bounded by the number of partitions $P$.  If we assume that we can store the {\tt pointer-array} in memory using Elias-Gamma encoding, we need to do one disk access per edge-partition that has any out-edges for the query vertex. In the worst case, each out-edge of the vertex is in a different partition, so the number of random accesses is bounded by the minimum of $P$ and the out-degree of the vertex. In addition, we need a sufficient number of sequential reads to transfer the edges to memory. The cost of the out-edge query is then:
\[
	\mbox{ io-cost[outquery(v)]  } \le \min(P, \mbox{outdeg(v)}) + \lfloor \frac{\mbox{outdeg(v)}} {B}\rfloor
\]

If we cannot keep the pointer-arrays in RAM, the first part of the bound is $2P$.   Compared to a standard adjacency list stored as CSR, the PAL structure has read-amplification of the order of $P$. On very large graphs, $P$ is typically in the hundreds, so the cost is significant. This trade-off buys us improved write throughput (Sec. \ref{sec:LSM}) and improves the locality of in-edge access, discussed next.  
Importantly, compared to linked list structure used by Neo4j, the number of random accesses is ultimately bounded by $P$, not by the size of the vertex neighborhood. Also, the out-edge query can be efficiently parallelized by querying each of the $P$ partitions simultaneously. PAL is indeed designed for modern SSDs that offer fast random access and multi-threaded access.





\subsubsection{In-edges}
\label{sec:inedges}

{\it In-edge query:  Given $u \in V$, and an edge $type$, return all $ \, (src, dst, \tau) \in E \,  | \, dst = u,  \tau=type$. In the social network example, for a query $u=$ `Jack', return all links \emph{from} the users that follow Jack.}

\smallskip

To find all the in-edges of a vertex, we need only to look at the one edge-partition that corresponds to the vertex-interval that includes the vertex ID. Inside that partition, the in-edges of the query vertex are in arbitrary positions, since the edges are stored in sorted order by the $source$ ID.   To avoid scanning the whole partition, we augment the CSR format by \emph{linking} all in-edges of a vertex together (unlike Neo4j, we use a single-linked list): as described in Sec. \ref{sec:structure}, each entry in the {\tt edge-array} contains offset to the next edge with the same $destination$ ID, or a special stop-word. In order to find the first edge in the chain, we do a binary search to the {\tt in-start-index} file. To limit the range of the binary search on disk, a sparse index can be used.   For each edge in the linked list, we search the {\tt pointer-array} to obtain the $source$ ID of the edge.

\mypar{Cost analysis: } To read all the in-edges we first issue one lookup to the {\tt in-start-index} with I/O cost of 1 (we assume a memory-resident sparse index that is consulted to find the file block containing the entry) and then traverse over the linked in-edges. In the worst case, each edge is on different block, requiring one access per edge, bound by the number of blocks in partition, which is on average $\frac{|E| }{PB}$ (average partition has $E/P$ edges). For each edge visited we consult the {\tt pointer-array}, which we assume (realistically) to be memory-resident due to the compression described previously. The total cost is then:
\[
	\mbox{ io-cost[inquery(v)]  } \le 1 + \min\Big(\mbox{indeg(v)}, \frac{E}{PB}\Big)
\]

If even the compressed {\tt pointer-array} does not fit into RAM, the I/O cost is doubled. Increasing the parameter $P$ decreases the size of the edge partitions, improving the locality of the in-edge query.  As with the out-edges query, the upper bound on the number of random accesses depends on $P$, not the size of the neighborhood. This property makes the PAL model attractive for power-law graphs which contain vertices of massive in-degrees. 

\mypar{Observation: } Increasing the number of edge-partitions $P$  has opposite effect on the random access cost of out- versus in-edge queries: out-edge queries become more expensive while the in-edge queries become less expensive. Whether searching for in- or out-edges is faster depends also on the in/out-degree of the query vertex (see Experiments). 



%
%
%
%

 \subsection{Edge  Data}

\begin{figure}[h]
\centering
   \includegraphics[width=0.4\textwidth]{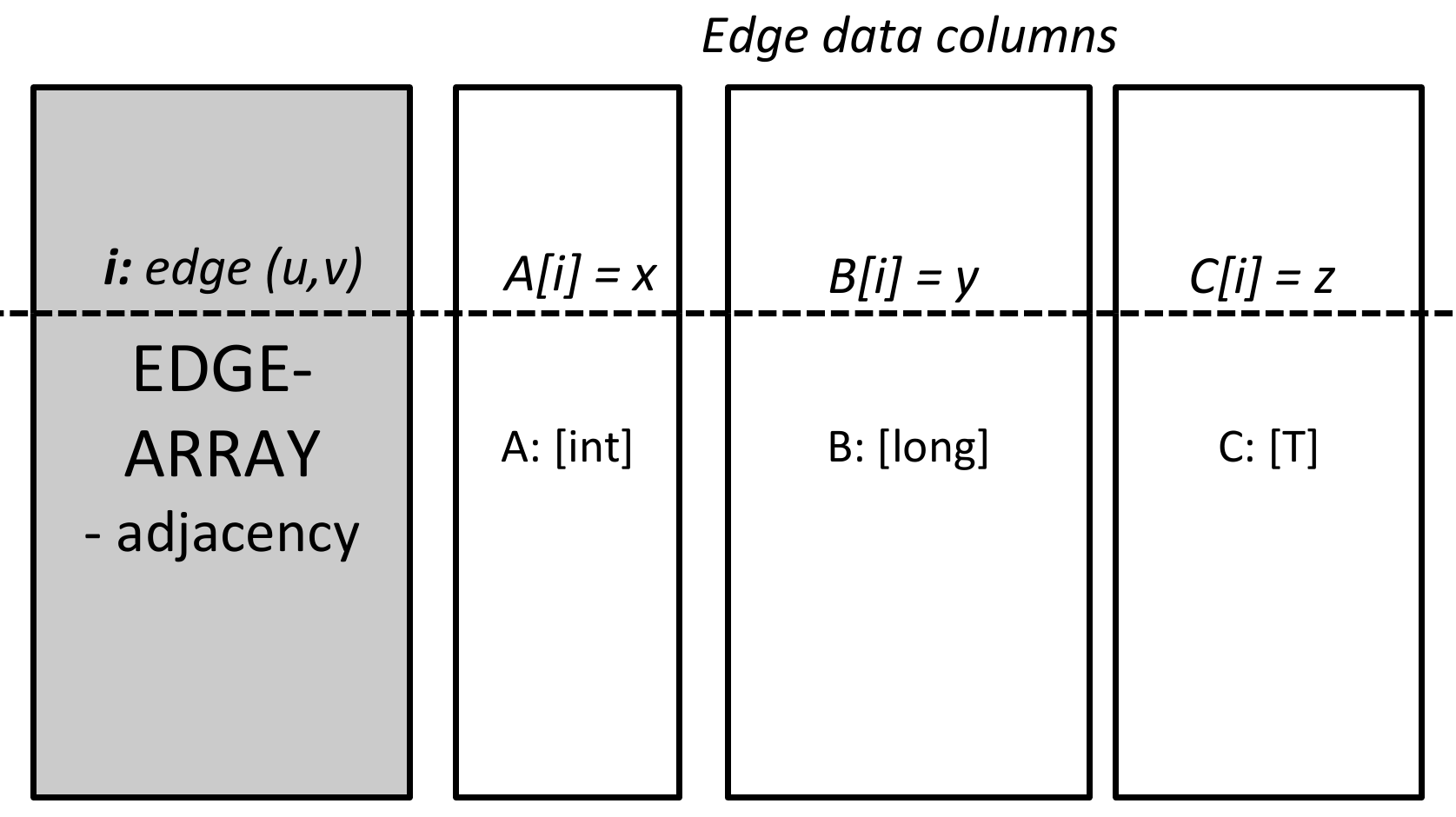}
\vspace{-4mm}
 \ncaption{ Column-oriented storage of edge data. For edge at position i in the edge-array, its attributes (x, y, z) can be found at index i from the columns A, B, C. }
 \label{fig:columnar}
 
 \end{figure}

The PAL model allows both row-oriented and column-oriented storage  \cite{stonebraker2005c} of the edge attributes. In the row-wise model, we would store the values  directly in the {\tt edge-array}, co-located with the adjacency information of the edge. However, we chose to implement columnar model for GraphChi-DB  because of its flexibility: we can add and remove columns without recreating the edge partitions.  Figure \ref{fig:columnar} illustrates the storage model. Each edge partition contains one file for each column (or ``field"') of an edge. These column-files are symmetric with the {\tt edge-array} file of the partition. If dense storage is used (i.e., each edge has a value set in the column), the column file is stored as flat array A where  value $A[i]$ has the value of edge at position $i$ in the {\tt edge-array}. Similarly, if the column values are sparse, for each non-null value we store a key-value pair where the key is the edge's position in the partition\footnote{Implementing sparse storage is future work.}.  Note  that if some fields are often accessed together, the columns can be combined. 

This model is very simple and efficient. Unlike many other solutions (such as DEX \cite{dex}), PAL does not require an additional foreign key to access the edge attributes. Instead, the position of the edge in the {\tt edge-array} is used to locate the corresponding attributes in the column files.   Note that this model also allows for looking up an edge quickly based on attributes: for a value matching a criteria in column at position $j$, we can efficiently find the edge object $(src, dst, \tau)$: $dst$ and $\tau$ are stored in {\tt edge-array[j]} and the {\tt pointer-array} is searched (similarly as when retrieving in-edges) for $src$.   Attributes can be also indexed using standard indexing techniques such as B-trees or bitmap indices. 

\subsection{Vertices}

Our solution to store the vertex data is extremely simple. We are only concerned with storing vertex attributes as the graph connectivity is fully stored in the edge partitions. 

Vertex data is stored exactly like edge data, in column files. The column files are partitioned based on the partition interval of the vertex. To access value of vertex $v$ from column $C$, we first find the vertex interval that includes $v$ and compute its offset from the start of the interval. That is, if $v=260,379$ and it belongs to the interval $p$ = [250,000 -- 500,000], the offset is $10,379$.  Then, the value is stored at position $10,379$ in the partition $p$ of column $C$.  Vertex value accesses have thus cost of only one I/O. However, if the vertex IDs are distributed very sparsely, using a sparse storage would be more efficient, leading to logarithmic access time but much smaller data.   


\section{Fast Edge Insertions}

This section describes how we allow online inserts of new edges to the database, with very high throughput. We start with the basic idea and then show how to make it scalable. The insert throughput of a graph database is crucial, because many large graphs also grow rapidly. 

\label{sec:inserts}

\subsection{Edge Buffers}

 \begin{figure}[h]
 \centering
   \includegraphics[width=0.25\textwidth]{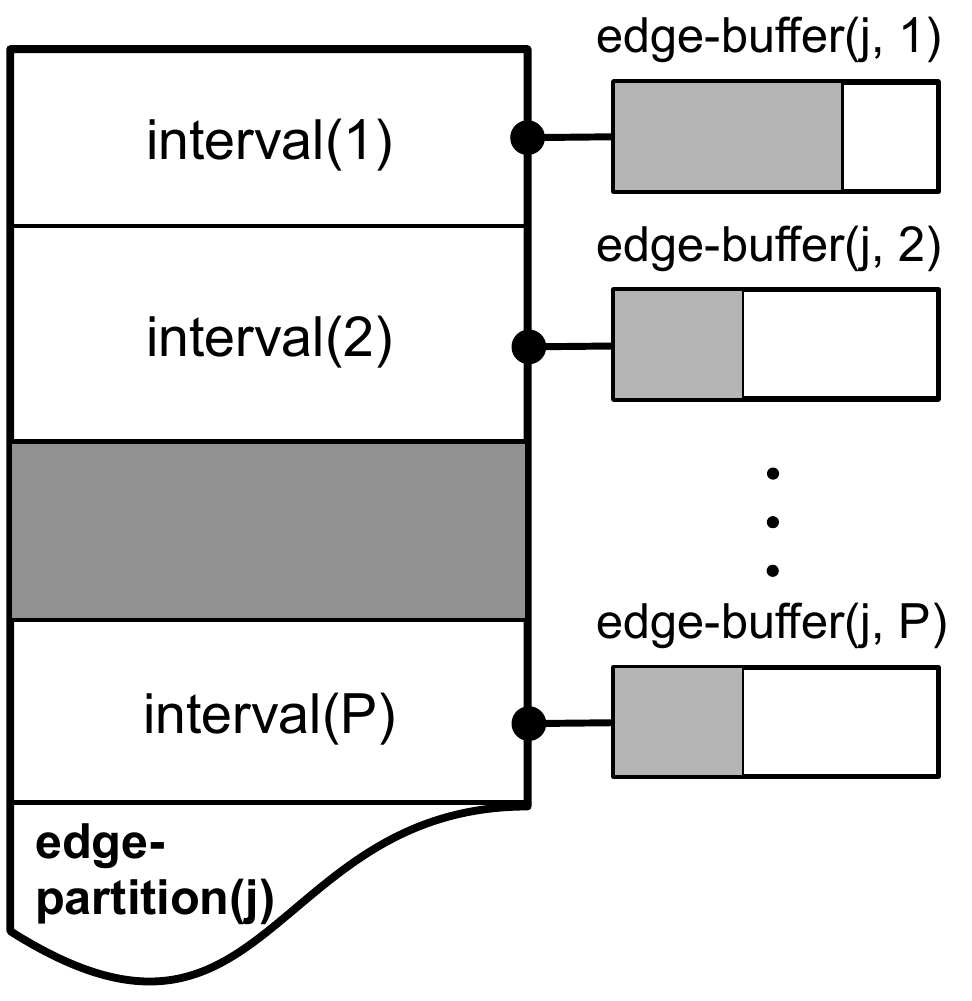}
   \vspace{-3mm}
\ncaption{ Edge buffers. Each edge partition can be split logically into $P$ subparts corresp. to the vertex intervals. Each subpart is associated with an edge-buffer. }
  \label{fig:edgebuffers}
 
 \end{figure}

Edge partitions are immutable data structures, so we cannot insert edges directly into them. Instead, new edges are first collected into buffers which are then merged in bulk with existing edges on disk to create new edge partitions. Following basic design was proposed in \cite{graphchi:osdi2012}. We split each edge partition logically to $P$ parts and attach each with an in-memory edge buffer, see Figure \ref{fig:edgebuffers}. New edges are immediately added to the corresponding buffer, based on the $source$ and $destination$ IDs. When we execute edge queries or computation (see Sec. \ref{sec:psw}), the buffers are also searched.  Note that buffers also store edge columns (attributes). 

When the number of buffered edges exceeds a threshold,  the buffer with most edges is merged with the edges on disk to create a new partition. This requires the old partition to be read from disk, sorting of the new edges in memory and a merge. In addition, creating the in-edge links described in Sec. \ref{sec:inedges} we need to sort the edges once. The IO-cost of the merge consists of reading the old partition and writing of the new partition.  If a partition becomes too large to fit in memory, it can be split it into two (see \cite{graphchi:osdi2012} for illustration).

While this technique works well initially, it unfortunately becomes inefficient when the graph grows. Consider that when a new partition is created the previous partition needs to be completely read from the disk and written back with the newly merged edges. This means that the first edges of the graph are rewritten as many times as the buffers are flushed. If the threshold size of buffers is $R$, then some edges may have been rewritten $E(t) / R$ times, where $E(t)$ is the size of the graph at time $t$. Another way to see the problem is to consider that the size of data on disk is, say, $100$ GB and the buffers can hold maximum $1$ GB worth of edges. Then the ratio of new edges to old edges is just $\frac{1}{100}$. 
 
\subsection{Log-Structured Merge-tree (LSM-tree)}
\label{sec:LSM}

\begin{figure}[h]
 \centering
   \includegraphics[width=0.48\textwidth]{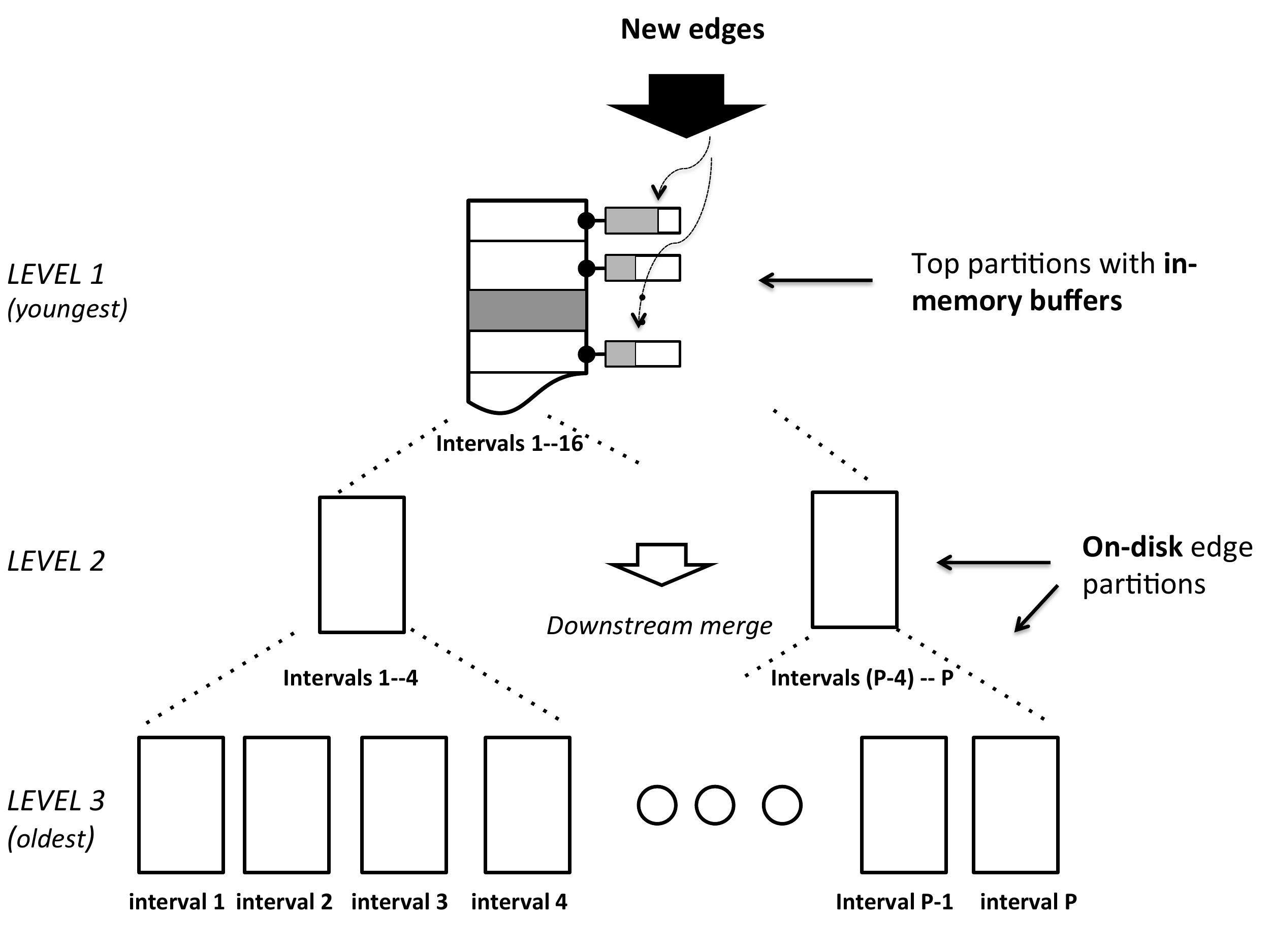}
\ncaption{Edge partitions in a LSM-tree. }
  \label{fig:lsm}
 
 \end{figure}

Our solution to the scalability problem with edge buffers is based on the Log-Structured Merge-tree \cite{LSM}, which is a write-optimized data structure based on stacking B-trees into an overlay tree structure.  In our case, we stack edge-partitions instead, as shown in Figure \ref{fig:lsm}. At the bottom of the tree we have the original $P$ edge partitions. The next level contains $\frac{P}{f}$ partitions, where $f$ is a branching factor (we use  $f=4$ in our experiments). Each leaf partition is associated with one vertex interval, as defined earlier. But an internal partition is associated with the \emph{union} of the intervals of its children. For example, in Fig \ref{fig:lsm}, the left-most partition on level 2 can store edges with $destination$ IDs in intervals 1 to 4. Otherwise, internal edge partitions are functionally no different from the original partitions and have exactly the same structure.  Note that each partition in the tree also includes the edge data column partitions.

Only top-level partitions (we assume existence of an imaginary root for the tree, not shown in the figure) have in-memory edge-buffers. When new edges are inserted, they are placed into the appropriate edge buffers of the top-level partitions. When the number of buffered edges exceeds a threshold, buffered edges are merged with the top-level partitions on disk. Furthermore, if a top-level on-disk partition exceeds a given size threshold, it it emptied and all its edges are merged to its child partitions. Similarly when an internal partition fills up, it is merged downstream to its child partitions. If leaves grow too large, we can add a new level into the tree.

The LSM-tree ensures that each edge is only rewritten a logarithmic number of times as function of the graph size, compared to linear times in the basic model, greatly enhancing the write throughput of the system. 
Note that $f$ can be adjusted based on the target workload (actually, the tree is not even required to be symmetric).

\subsubsection{Trade-off with Query Performance}

The improved write-throughput of LSM comes of course with a trade-off with regards to read performance. We review the effects below. For analysis, we introduce following notation:  $ L_G := $ the number of levels of the LSM-tree; $ P_G(j) :=$ number of partitions on level $j   \in {1,...,L(G)}$.

\mypar{Out-edge queries: }  To find the out-edges for a query vertex $v$, the database needs to look at all the partitions on all the LSM-tree levels:
\[
	\mbox{ io-cost[outq(v)]  } \le \min \Big(2\sum_{i=1}^{L_G} P(i), \mbox{outdeg(v)}\Big) + \lfloor \frac{\mbox{outdeg(v)}} {B}\rfloor
\] 
\mypar{In-edge queries: }  The in-edges of a vertex $v$ can now be found from any of the partitions associated with a vertex interval that contains $v$. Fortunately, we can look at those partitions in parallel. There is only one such partition on each level, so the cost is bounded by:
\[
	\mbox{ io-cost[inq(v)]  } \le L_G + \min\Big(\mbox{indeg(v)},  \mbox{max-partition-size}/B \Big)
\]

 Above, we introduced a parameter {\it max-partition-size} bounding the size of any partition in the LSM-tree. 

%

We believe the additional cost in read time is acceptable (on SSD), because the number of total edge partitions increases at most by two (with $f=2$). 
Also, the structure is very flexible and the user of the database can adjust the thresholds of the edge-buffers and modify the structure of the LSM-tree
to match a desired target workload. 



\vspace{-1mm}

\subsection{Updating Edge Attributes and Deletes}

The LSM-tree model could also be used for updating edge attributes: to modify an edge, we would first read its current values and then insert a new edge with the updated values into an edge-buffer. When the edge would be eventually merged with a partition storing an older version of the edge, the older version would be removed. A similar approach can be used for deleting edges. This would enable similar throughput of updates as for inserts, but the downside is that it would make queries slightly more complicated as we would need to check for duplicate edges and choose the newest one. A bigger problem would arise with analytical computation that streams over edges: before handling a streamed edge we would need to ensure that a more recent edge was not processed before.  

To simplify the design of GraphChi-DB,  we instead implemented updates and deletes by directly modifying the edge attributes in the column files of the edge partitions. This solution is acceptable if the frequency of updates or deletes is relatively small.  Edge deletions are handled by \emph{tombstones}: permanent removals take effect during partition merges.


%
%

\section{Graph Computation}

\label{sec:psw}

\begin{figure*}[t]
\includegraphics[width=0.95\textwidth,clip]{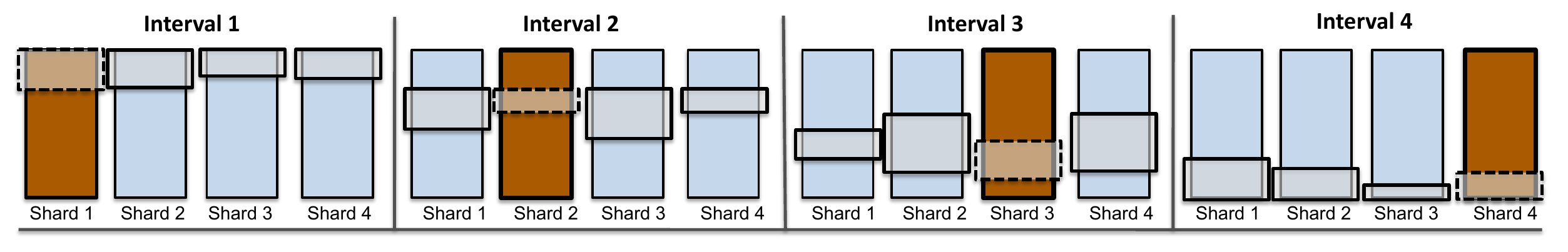}
\vspace{-4mm}

\ncaption{The phases of one iteration of the PSW algorithm. 
In this example, vertices are divided into four intervals. The computation proceeds by constructing a subgraph of vertices in one interval
a time. In-edges for the vertices are read from the {\bf owner partition} (in dark color) while out-edges
are read from the {\bf ``sliding windows"}. } \label{fig:phases}
\end{figure*}
 
As the PAL structure is based on the design introduced in \cite{graphchi:osdi2012} for GraphChi, it can also be used for efficient analytical computations.  We describe this functionality only briefly. 

\subsection{Parallel Sliding Windows (PSW)}

PSW is based on the vertex-centric model of computation popularized by Pregel \cite{pregel} and GraphLab \cite{uaigraphlab}. In that model, programmer specifies an {\bf update-function} that is executed for each vertex in turn. In addition to the vertex attributes, it  can access all the incident edges, their attributes, and modify them. Typical update function is presented in Algorithm \ref{alg:updatefunction}.

\begin{algorithm}[h]
   \ncaption{Skeleton of a vertex {\bf update-function}}
   Update(vertex) \Begin {  
   	x[] $\leftarrow$ read values of in- and out-edges of vertex \;
	vertex.value $\leftarrow$ f(x[]) \;
	\ForEach{edge of vertex} {		edge.value $\leftarrow$ g(vertex.value, edge.value)\;
	}
  }
  \label{alg:updatefunction}
\end{algorithm}

  \begin{algorithm}[h]
\ncaption{Parallel Sliding Windows}
\label{alg:psw}
 PSW(G, updateFunc) \Begin {
   \ForEach{interval $I_i \subset V$} {
   	$G_i$ := LoadSubgraph($I_i$) \\
	\ForEach{v $\in$ $G_i.V$} {
	     updateFunc($v, G_i.E[v]$) 
	}
	UpdateToDisk($G_i$) 
   }
}
\end{algorithm} 

PSW executes the update-function for all vertices of the graph in order. PSW 
processes graphs in three stages: it 1) loads a subgraph from disk; 2) updates the vertices and edges; and 3)
writes the updated values to disk.  This process is repeated for all subgraphs until the whole graph is processed.
The subgraphs correspond to the vertex-intervals defined earlier:  subgraph $i$ contains the incident edges of vertices of interval $i$. 
Pseudo-code is shown in Alg. \ref{alg:psw}. 

The PSW algorithm is efficient because reading and writing of a subgraph requires only a small number of random accesses: to load the in-edges for the vertices, the edge partition corresponding to the subgraph interval is loaded completely in memory (thus, we require partitions to be small enough to fit in RAM); the out-edges are then in consecutive ``windows'' in all the of the partitions. Thus, only $\Theta(P^2)$ random seeks are required to process the whole graph. See Figure \ref{fig:phases} for illustration and we refer to \cite{graphchi:osdi2012} for more details.

{\it Cost analysis:}  The I/O-cost of one iteration of PSW was derived in \cite{graphchi:osdi2012},  but we modify it to for the LSM-tree:
\[
  \frac{2|E|}{B} \le PSW_{B}(E) \le \frac{4|E|}{B} + \Theta(\Big[\sum_{i=1}^{L_G}P(i)\Big]^2)
\]
\newpage
\subsubsection{Alternative Models of Computation }

In the PSW model, the state of the computation is encapsulated in  the edge values. It also allows random access to the whole neighborhood of a vertex in an update-function. However, many algorithms such as Pagerank \cite{pagerank} store state in the vertex values (typically $|V| << |E|$) and do not require random access to the neighborhood. Fortunately, GraphChi-DB can also support the edge-centric model such as in X-Stream \cite{xstream} or GraphLab \cite{powergraph}. In that case, a typical computation proceeds by streaming in-edges of an edge partition and executing a function for each of the edges, typically   reading value of the $source$ vertex of the edge and changing a attribute of the $destination$ vertex of the edge. This requires storing $O(V)$ values in memory, which is often feasible. If memory capacity is limited, it is also possible to execute the computation in multiple sweeps so that only a subset of vertices is accessed at any time (see \cite{xstream}). 

\subsubsection{Incremental Computation}

Many analytical computations, such as Pagerank and label propagation algorithms, are based on fixed-point iteration. As the PAL allows execution of computation `in-place', we can execute graph programs \emph{incrementally}, similarly to Kineograph \cite{kineograph}   in the distributed setting. In the incremental setting computation is triggered by changes in the graph, and continuous vertex updates are performed in order to keep the computational state, such as an authority-score of a user vertex, as up-to-date as possible. In contrast to Kineograph which defines computation on a fixed snapshot, we allow the graph to continuously change (although we could define logical snapshots based on timestamps). Thus, the computational state may never match the current state of the graph, but we argue that for many cases this can be tolerated.   We caution that all types of computation are not amenable for incremental computation: for example, the connected components of a graph can drastically change after an edge is removed, and it is not clear how to update the component labels without recomputing from scratch.

Due to lack of space, we do not discuss incremental computation further. Many research questions also remain open: for example, what kind of computation is valid to execute incrementally, and how to analyze the error of the current computational state when the graph is constantly changing.


%
%
\newpage
\section{GraphChi-DB: Implementation}

Several details need to be addressed by a database implementation based on the Parallel Adjacency Lists structure. We engineered GraphChi-DB mostly with the Scala programming language \cite{odersky2004overview} that runs on the Java Virtual Machine.  In addition, small parts are written in Java and most frequently used sorting routines are written in C++ for maximum performance. GraphChi-DB is an embedded database management system. 

\subsection{Buffer Management}

GraphChi-DB is optimized for data that is much larger than the available memory. Thus, only parts of the graph are at any time paged in RAM. To greatly simply the engineering effort, we used  the \emph{memory mapping} functionality provided by the operating system, through Java's interface. (Memory mapping was previously used for disk-based graph computation in \cite{sabrin2013mmap}). Memory mapping uses the virtual memory system so files can be accessed through ordinary pointers. The OS takes care of transferring the data between disk and memory transparently.  In our experience, the memory mapping provided sufficient performance and greatly simplified the code.

\subsection{Vertex IDs and Intervals}

\label{sec:vertexid}

Recall that edges are divided by their $destination$ ID to the edge partitions, based on which vertex interval contains the ID. In most cases the edges are not distributed evenly into the partitions and some partitions could be many times larger than others, requiring us to dynamically manage the vertex intervals for balance. This solution is adopted in \cite{graphchi:osdi2012}, but for GraphChi-DB we decided to use a simpler solution, based on using a reversible hash function.  We split the range of vertex IDs  $0..N$ into $P$ equal length vertex-intervals of length $L$ and map each original ID into an {\bf internal ID} (and back) as follows:

\smallskip

\begin{small}
\begin{tabular}{p{0.95cm}  p{0.1cm}  l } 
{\tt intern-ID } & {\tt  := } & {\tt  (orig-ID mod P) L + (orig-ID $\div$ P)} \\
{\tt orig-ID} & {\tt :=  } & {\tt  (intern-ID $\div$ L)P + (intern-ID mod  L) }  \\
\end{tabular}
\end{small}

\vspace{-3mm}

Assuming the distribution of the edges does not correlate strongly with the $P$-modulo of the $destination$ ID, edge partitions will be sufficiently balanced in practice. This solution is not adversary-proof but works well in practice and simplifies the implementation significantly. Fixed-length vertex intervals provide also performance advantages as we can find the interval for a vertex ID mathematically. 

\subsection{Consistency}

GraphChi-DB provides ``fire-and-forget" semantics for transactions: writes are immediately visible to concurrent sessions (we did not implement higher level transactions as we argue they are rarely needed in this context). Depending on the reliability requirements, GraphChi-DB can operate in one of two modes: with durable buffers and only-memory buffers. In the latter case new edges are added directly to the edge-buffers and do not survive a crash that would happen before the buffer is merged with the edge partitions on disk. With durable buffers, each edge insertion is first written into a log-file that is synced with the filesystem to guarantee durability over computer crash. The durable option has significant performance cost, but as it is constant per edge, it does not affect the scalability properties of the system. 

GraphChi-DB's transaction processing is also greatly simplified by the PAL's flat data structures: integrity of the data structure itself cannot get compromised by a hardware failure. When a new partition is created from merged edges, the old partitions are discarded only after the new partitions have been successfully committed to disk.
 

%
%
%
\vspace{-1mm}

\subsection{Queries}
\label{sec:queries}

PAL-model would work with any graph query language such as Cypher \cite{neo4j} or SPARQL\footnote{http://www.w3.org/TR/rdf-sparql-query/}. 
In this work we concentrated on the data structure and its properties, and did not implement a query language for GraphChi-DB, but instead provide a Scala-API to access the graph.  As an example, below is actual code for finding the ``friends-of-friends'' of a user, excluding the friends themselves:
\vspace{-1mm}
\begin{footnotesize}
\begin{verbatim}
 val friends = queryVertex(queryId, DB)
 val result = friends->traverseOut(FRIENDS)
     ->traverseOut(FRIENDS)
     ->selectOut(FRIENDS, 
      dst => !friends.hasVertex(dst))

\end{verbatim} 
     \end{footnotesize}
The traversal operator {\tt traverseOut} visits all out-edges of the vertices in the current {\it frontier} (set of vertices) and adds their \\$destination$ IDs to the next frontier. 
We use a simple optimization proposed in \cite{beamer2013direction}: if the current frontier is very large, to compute the next frontier, instead of issuing a (top-down) out-edge query for each of
the vertices in the frontier, it can be more efficient to (bottom-up) sweep over \emph{all} edges of the graph and for each edge check if they are in the current frontier. Our traversal operators are similar to the operators of Ligra \cite{shun2013ligra}, a system for in-memory graph computation. Thanks to the Scala's powerful syntax and the REPL (Scala console), we believe the Scala API to be good alternative for a query language.

%

\section{Experiments}

\label{sec:experiments}
%
%
%

In this section we put our design to test and also compare against existing systems. We start with an experiment that simulates a challenging online graph database workload, demonstrating the scalability of GraphChi-DB. We then evaluate how the LSM-tree improves the write performance and compare to Neo4j. Finally, we study GraphChi-DB's performance in serving graph queries and evaluate different choices in indexing for the edge partitions.  
  For experiments we used two machines:  (1) Mac Mini with dual-core Intel i5 CPU, 8 GB of RAM and a 256 GB SSD; (2) MacBook Pro laptop with 4-core Intel i7 CPU, 8GB of RAM and a 512 GB SSD.

\subsection{Database Size}
 \begin{table}[hbt]
\begin{small}
\begin{tabular}{| l | c | c | c | c | }
\hline
Graph & Edges & Chi-DB & Neo4J & MySQL \\
         &     &                      &           & data + indices \\
\hline
live-journal \cite{livejournal} & 69M & 0.8 GB & 2.3 GB & 2.1 GB \\
twitter-2010 \cite{kwak2010twitter} & 1.5B & 17 GB  & 52 GB & -- \\
LinkBench  & 5B &  $\sim$ 350GB & -- & 1400 GB \cite{linkbench} \\
\hline
\end{tabular}
\end{small}
\vspace{-3mm}
\ncaption{Comparison of database disk space for graphs stored in GraphChi-DB, Neo4j and MySQL. The size of the LinkBench graph for GraphChi-DB has been estimated upwards to account for the slightly the smaller edge attributes than used for MySQL (see text).}
\label{table:dbsizes}
\end{table}

Table \ref{table:dbsizes} displays the size of the database files with different database management systems. 
Neo4j uses exactly 35 bytes for each edge, compared to 9 bytes by MySQL (using 4-byte integers for vertex IDs; if 8-byte integers would be used, the size of the data would double). However, for the relational database MySQL, the indices over the data are much larger than the data itself. GraphChi-DB has the smallest database size, but the number of bytes required per edge varies slightly depending on the number of edge partitions used and the distribution of edges.

\subsection{Online Database Benchmark: LinkBench}

Our first set of experiments is based on a graph database benchmark LinkBench published by Facebook \cite{linkbench}. LinkBench includes a generator that generates a graph that has similar structural properties as Facebook's social graph. It then performs a highly parallel request workload with a mix of inserts, updates, deletes and immediate neighborhood queries. Since LinkBench only queries the first level out-neighbors of nodes, it is a workload well suited for a relational SQL database, and does not stress the graph traversal features that most graph database management systems are optimized for (we evaluate traversal queries in Section \ref{sec:queryexp}). Another important feature of LinkBench is that it is designed for a workload that has no locality of access: that is, its workload simulates the requests to fulfill cache misses, assuming there is a large in-memory caching layer (see Facebook's Tao architecture \cite{venkataramani2012tao}) that absorbs frequently issued reads.  We note one short-coming of the LinkBench benchmark: the edges are not assigned randomly, but each vertex $u$ has edges to its neighbors with subsequent IDs $u+1, u+2, ...$. Thus, there is more locality in access than realistically.

The data model defined by LinkBench is as follows. Each edge and vertex has four fields: type, timestamp, version and a random payload string\footnote{GraphChi-DB stores variable length data by writing them into a separate log, similarly to a Log-Structured Filesystem \cite{rosenblum1992design}. The log-position of the value is then stored as the edge or vertex attribute. Due to lack of space, we do not discuss the implementation in more detail.}. The length of the payload string is random -- for details, see\cite{linkbench}. In GraphChi-DB's LinkBench implementation we ignore the vertex type (as there is only one type in the benchmark) and use 4 bits for edge type (benchmark uses only two edge types).  

We executed the benchmark on the MacBook Pro laptop, with 64 concurrent request threads (due to high amount of I/O, GraphChi-DB's throughput improved with higher concurrency, peaking at 64 threads). We did not use durable buffers. We configured the benchmark to generate 1B vertices, with approx. 5B edges. Summary of our results is shown in Table \ref{tab:linkbench}. We compare the results to those of Facebook, achieved on MySQL running on a high performance server with 144 GB of RAM and a SSD array. Unfortunately, the results are not directly comparable as we could not repeat Facebook's experiment ourselves\footnote{Also, we could not repeat the MySQL experiment on a laptop since the graph data requires about 1.4TB of space, \cite{linkbench}, dwarfing the capacity of our SSD.}, and did not have access to similar hardware. There is also small difference in the size of edge and vertex data used in the experiments (this was required to fit the graph on our SSD), but the impact should not be material.
That caveat in mind, the results show that GraphChi-DB can efficiently handle very large graphs with  a challenging mix of reads and writes on just a laptop.  We now discuss the results in more detail.

Results in Table \ref{tab:linkbench} show that all vertex-related operations are extremely fast on GraphChi-DB, because they require just constant time access due to the static indexing. On the other hand,  request {\tt edge\_insert-or-update} has high latency because it requires that if the edge to be written already exists, it must be updated (with a direct write to the edge partition, often resulting in a page miss). JVM's garbage collection pauses are the primary reason for the high 95-percentiles for GraphChi-DB. We also note that because requests {\tt edge\_getrange} (out-edges of a vertex in a timestamp range) and {\tt edge\_outnbrs} require sorting of the results by timestamp, the benchmark is also CPU-intensive. Removing the sorting from the benchmark increases GraphChi-DB throughput by about 20\%.  In general, GraphChi-DB has lower latencies for writes, but Facebook's system is faster with queries. 

\begin{figure*}[t]
\begin{center}
  \subfigure[LinkBench request throughput] {
   	\includegraphics[width=0.32\textwidth]{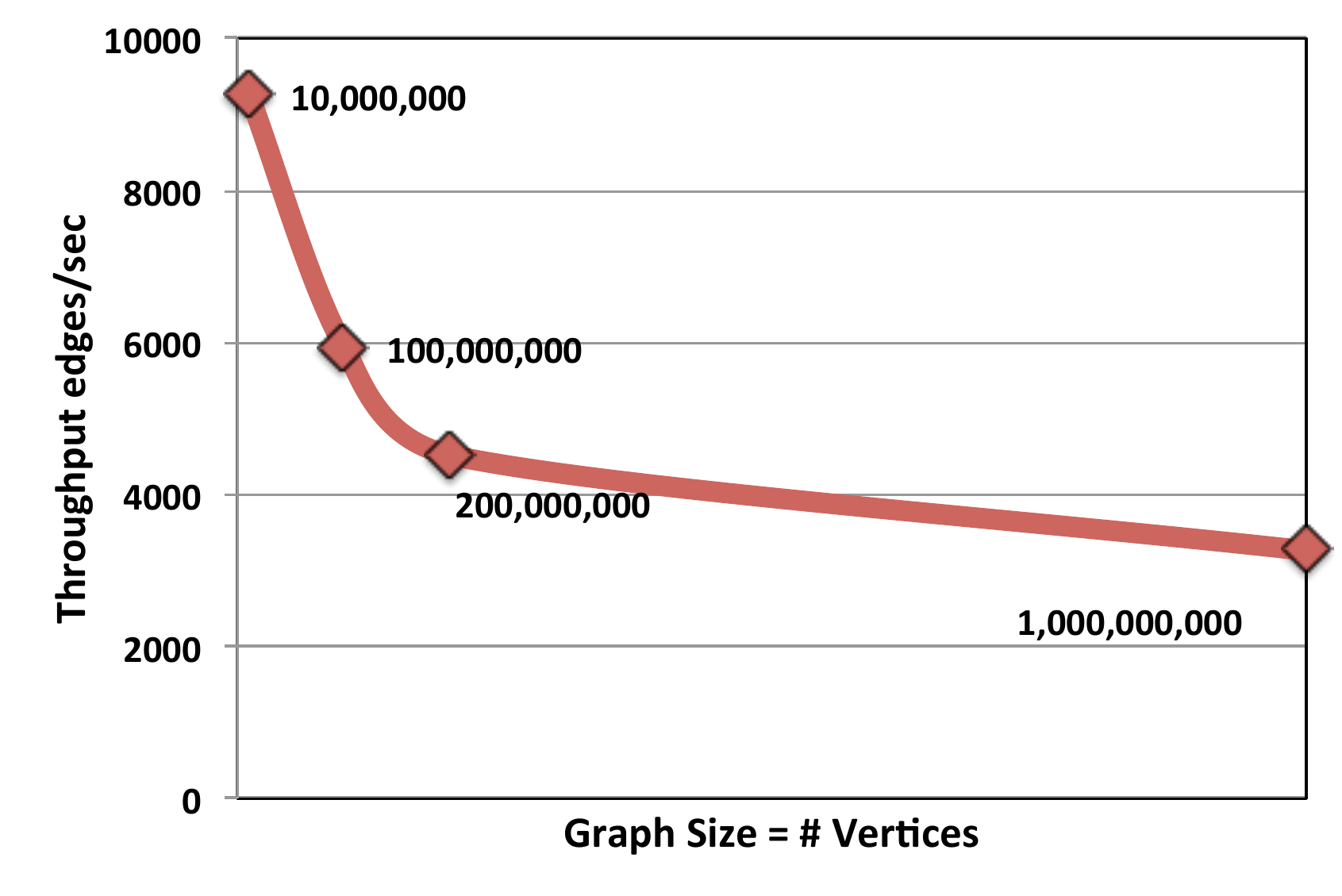}
  }
  \subfigure[Friends-of-Friends] {
 	\includegraphics[width=0.32\textwidth]{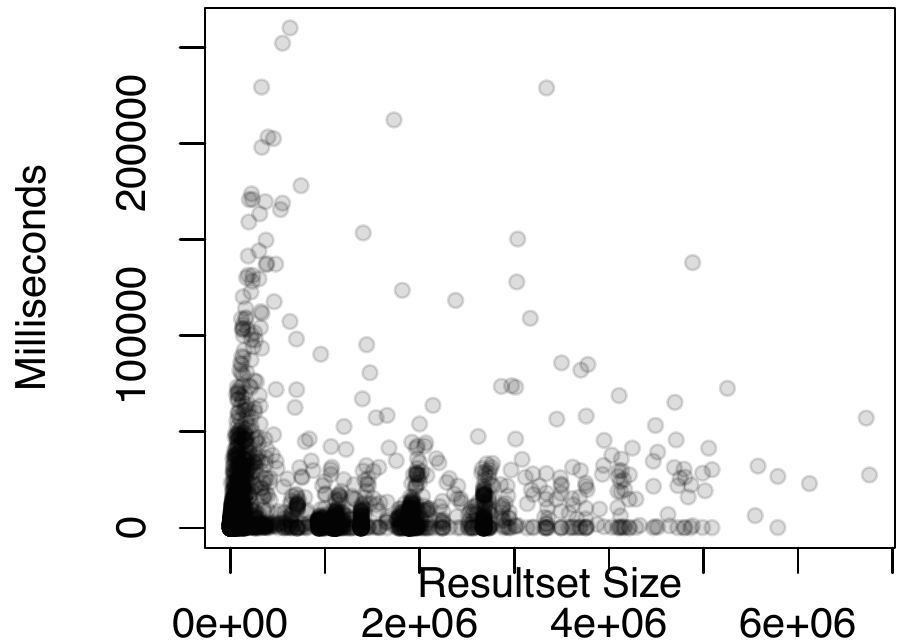}
   }
     \subfigure[Pointer-array indexing] {
 	\includegraphics[width=0.27\textwidth]{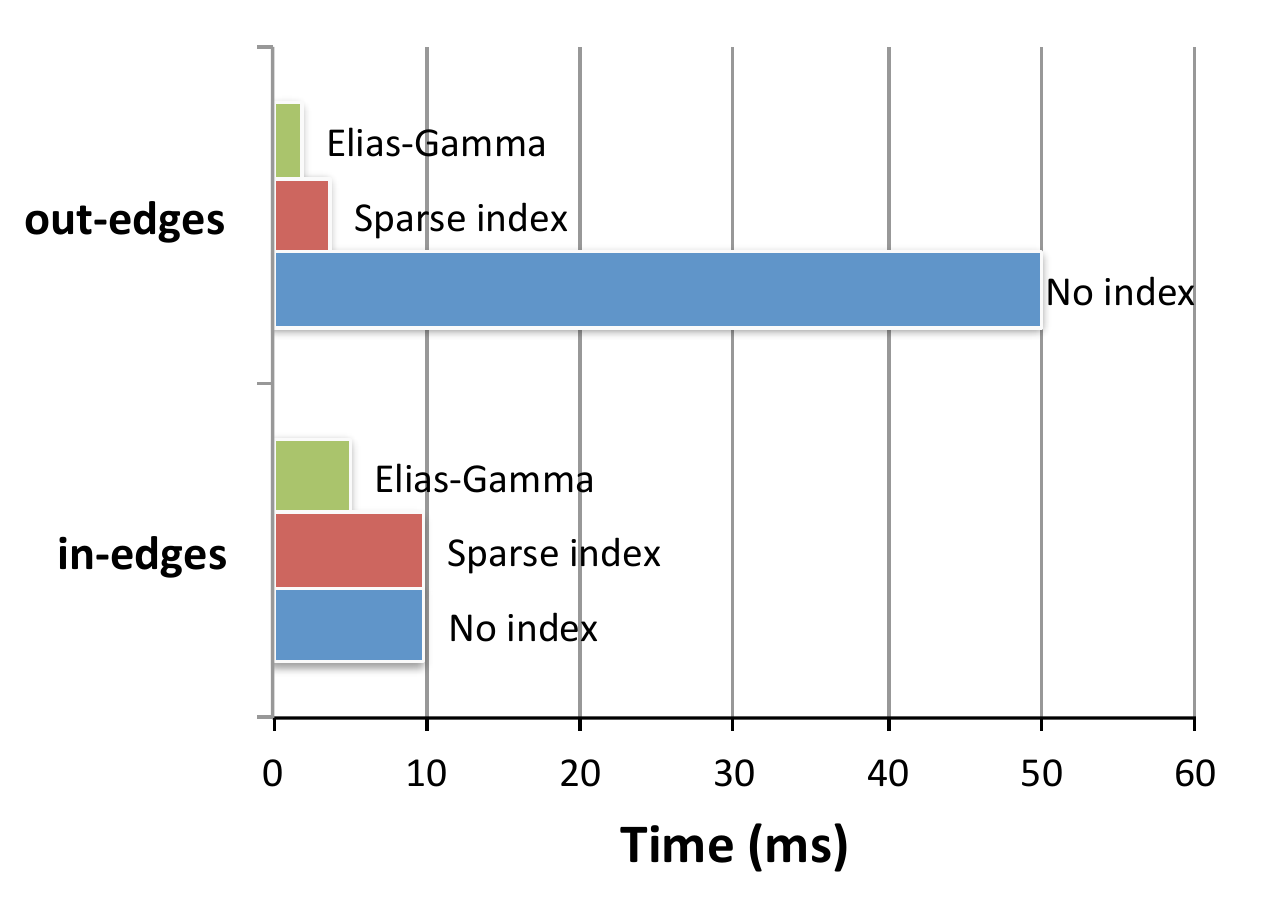}
   }
   \vspace{-4mm}
	\ncaption{(a) GraphChi-DB throughput on the LinkBench benchmark as function of graph size. Number of edges in the graph is approx. five times the number of vertices. (b) Friends-of-Friends query time distribution as function of result set size (Mac Mini) on the Twitter graph. Interestingly, the query time does not correlate much with the final result set size --  likely because the I/O access is dominated by the random seeks and thus the time to read edges of a friend does not vary much based on the number of edges. High variance is explained by the large size of the graph that requires constant paging of data from disk to RAM. (c) Impact of the choice of indexing for the pointer-array of an edge partition to in- and out-edge queries. Figures are averages over 80,000 runs.  }
	\label{fig:fof}
	\label{fig:linkbenchgsize}
	\label{fig:indexstudy}
	\end{center}
\end{figure*}

Figure \ref{fig:linkbenchgsize} shows how the throughput of GraphChi-DB scales with the size of the benchmark graph, validating the scalability of the system. Smaller graphs fit completely or largely in RAM, and thus show better performance. 

We also implemented a LinkBench driver for the commercial DEX \cite{dex} graph database. Unfortunately, we were not able to create very big graphs for DEX, as after the graph size exceeded DEX's cache size, the insertion rate stumbled to less than one thousand edges / sec  (we consulted with the developers of DEX in attempt to resolve the problem). However, we managed to run LinkBench on a small graph with 5M vertices and 25M edges. We run DEX in auto-commit mode, with recovery functionality disabled.  On the MacBook Pro, DEX could execute 1,000 requests / sec using 16 threads. Using the same 16 threads, GraphChi-DB outperformed DEX with a throughput of 7,900 edges / sec. 
 
 
 \begin{table}
 \begin{small}
 \begin{tabular}{| l | c | c | c || c | c | | c |}
 \hline
   & \multicolumn{3}{|l||} { GraphChi-DB}  &  \multicolumn{3}{|l|} { MySQL + FB patch }   \\
    & \multicolumn{3}{|l||} {\it laptop (SSD)}  &  \multicolumn{3}{|l|} {\it server (SSD-array) \cite{linkbench}}   \\
\hline
     &   p50 & p75 & p95 & 50 & p75 & p95 \\
   \hline
   node\_get      &2 & 4 & 34      & 0.6 & 1  & 9  \\
   node\_insert     &  0.1 & 0.1 & 0.1     & 3& 5 & 12 \\
   node\_update & 2  & 4 & 34          &  3 & 6& 14  \\
   edge\_ins-or-upd.       & 0.7 & 2 & 15  & 7 & 14 & 25  \\
   edge\_delete    & 0.1 &0.9 & 7    & 1 & 7 & 19 \\
   edge\_update   & 1 & 3 & 22 & 7 & 14  & 25 \\
   edge\_getrange & 8 & 19 & 250 & 1 & 1  & 10 \\
   edge\_outnbrs & 0.4 & 3 & 18 & 0.8 & 1 & 9 \\
   \hline
   Avg throughput & \multicolumn{3}{|c||} {2,487 req/s} & \multicolumn{3}{c||} {11,029 req/s} \\
   \hline
 \end{tabular}
\end{small}
\vspace{-1mm}
\ncaption{LinkBench online database benchmark. Latencies  are in milliseconds. Note: for clarity we have modified the request names from the original.  JVM's garbage collection pauses cause the high 95-percentiles. }
\label{tab:linkbench}
 \end{table}

\subsection{Insertion Performance}

\begin{figure*}[t]
\begin{center}
  \subfigure[Inserted edges / time] {
 	\includegraphics[width=0.40\textwidth]{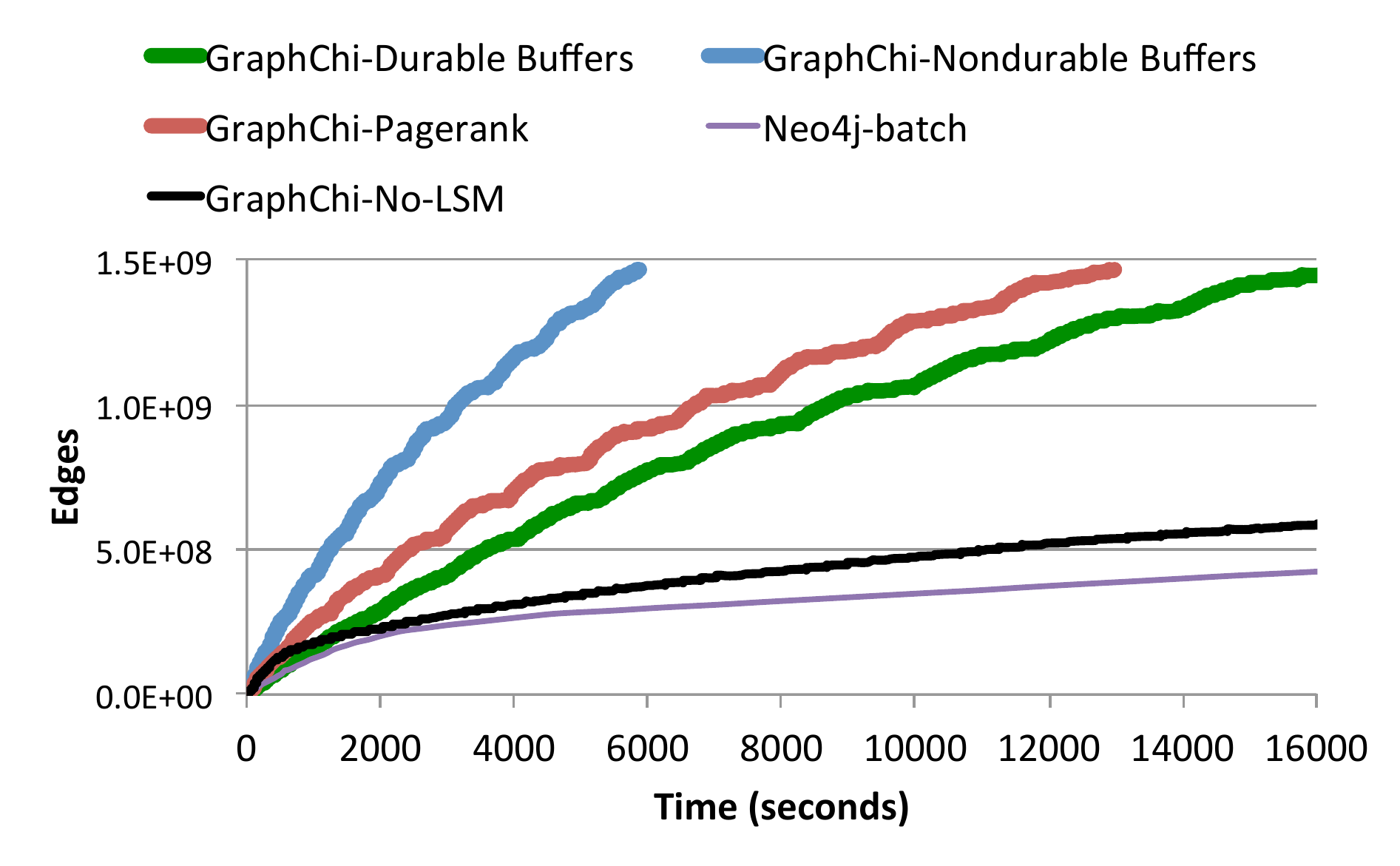}
  }
  \subfigure[In/out queries] {
 	\includegraphics[width=0.35\textwidth]{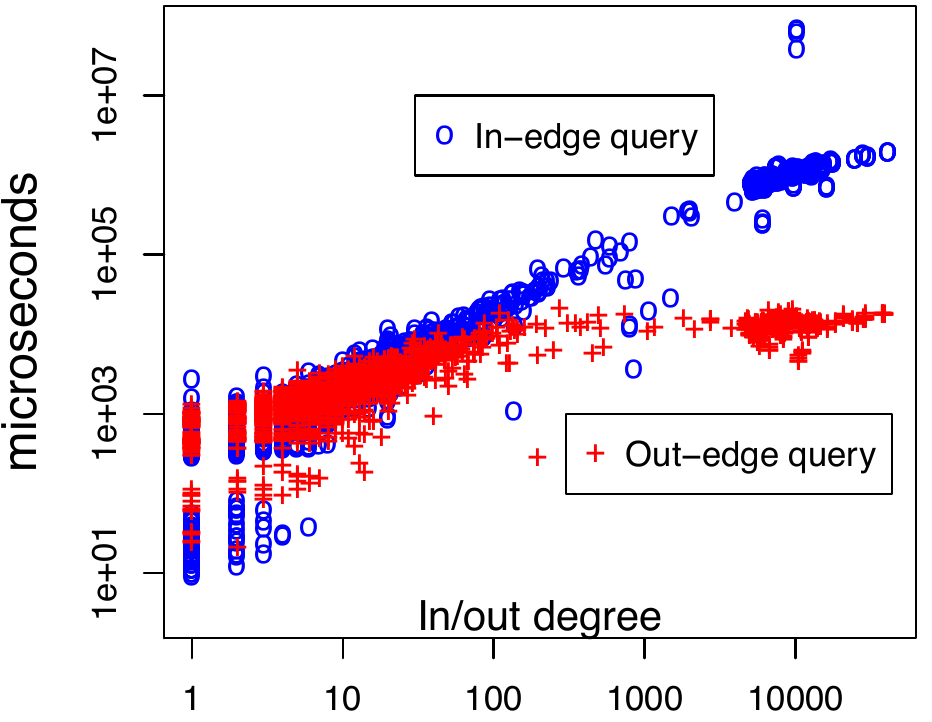}
  }
  \vspace{-5mm}
	\ncaption{(a) The number of edges added over time, on Mac Mini, {\it twitter-2010} graph. The first line shows the number of edges inserted over time by GraphChi-DB when buffers are not durable. Average insertion performance is about 248K edges/sec. Second line is for when GraphChi-DB executes Pagerank algorithm continuously while inserting. Third line shows the progress with durable buffers, and the black line is without LSM-tree (but non-durable buffers). Finally, the last curve shows Neo4j's batch importer's progress. It took about 45 hours to insert all 1.5B edges with Neo4J. Note that GraphChi-DB was run in  the online mode. (b) For the same graph, scatter plot of GraphChi-DB query time of in/out-edge queries for a random set of vertices. Both axis are in logarithmic scale. The time depends linearly on degree when the degree is small but retrieving large neighborhoods is relatively more efficient, which is in line with the theoretical analysis. }
	\label{fig:twitteringest}
	 \label{fig:inoutscatter}
\end{center}
\end{figure*}

Next we study the edge insertion performance of GraphChi-DB. Figure \ref{fig:twitteringest}a shows the number of edges inserted over time by GraphChi-DB and in comparison, Neo4j's batch inserter. If GraphChi-DB's edge buffers are not durable, the 1.5B edges of the {\it twitter-2010} graph an be inserted in about one and half hour (almost 250K edges/sec). Durable buffers add a constant cost to each edge insert, increasing the total to 4.5 hours. We also run the experiment without the LSM-tree, shown in black in the plot: insertion throughput drops quickly when the size of the graph increases.  We also run the insert experiment by executing Pagerank \cite{pagerank} algorithm simultaneously on the growing graph, demonstrating the ability of GraphChi-DB to run analytical computation on the data. In this case, the ingest took about 3.5 hours.  Note that we inserted edges in online mode to GraphChi-DB, while for Neo4j's we used their non-transactional batch importer. Still, it took over 45 hours to ingest the same data. 
These results demonstrate that on just a Mac Mini, GraphChi-DB can handle a extremely high rate of edges insertions even when the graph is very big.  

The wave-like pattern in the insert progress of GraphChi-DB is caused by stalls that happen when the buffers become full and the system is busy merging the buffers to the disk-based edge partitions. Write-stalls are inevitable because we inserted edges as fast as possible to the graph.


%
%
%
%

\begin{table}[t]
\begin{small}
\begin{tabular}{|l|c|c|c|c|}
\hline
 & \multicolumn{4}{c |}{ FoF query latency (ms) } \\

\hline
 {\bf livejournal} (1M queries) & 50p & 75p & 95p & 99p \\
\hline 
\hline
GraphChi-DB & 0.379 & 0.928 & 2.965 & 6.653 \\
Neo4J & 0.127 & 0.483 & 2.614 & 8.078 \\
{\it ratio} & 0.3$\times$ & 0.5$\times$ & 0.9$\times$ & 1.2$\times$ \\
  & & & & \\
GraphChi-DB+ Pagerank & 0.432 & 1.065 &  3.578 &8.233  \\

\hline
{\bf twitter-2010} (100K queries)   & 50p & 75p & 95p & 99p \\
\hline
\hline		
GraphChi-DB&	22.4 &	60.9 &	554.5 &	1,264 \\
Neo4J&	759.8 &	3,343&	26,601 &	86,558 \\
{\it ratio}	& 40$\times$ &55$\times$ & 48$\times$ & 68$\times$  \\
  & & & & \\
GraphChi-DB+ Pagerank & 28.1 & 76.3 &	705.8 & 	1,631  \\
	\hline
\end{tabular}
\end{small}
\vspace{-1mm}
\ncaption{  Friends-of-Friends Query latency quantiles on the Mac Mini. On the {\it livejournal} graph which fits in RAM, Neo4J services the queries roughly twice as fast as GraphChi-DB (with LSM-tree (16,4)), although the 99-percentile is slightly better for GraphChi-DB. But on the much larger {\it twitter-2010} graph,  GraphChi-DB outperforms Neo4j by a wide margin.  } 
\label{table:fof}
\end{table}

 \subsection{Graph Queries}

\label{sec:queryexp}

Next we studied the query performance of GraphChi-DB. In these tests, the graph was not modified.  In addition to the simple in- and out-neighborhood queries we evaluated two more challenging types of graph queries. 

\mypar{In- and out-edges: }  With PAL, the access of in- and out-edges is very different, so it is important to know the implications on the access latency (please note that we could switch the treatment of in/out edges).  Figure \ref{fig:inoutscatter}b shows the distribution of query times of both in- and out-edge queries as function of the vertex degree (i.e number of edges). For small results, in-edge queries are slightly faster, but if the number of edges exceeds 100, the out-queries are faster. In line with our analysis, for very large degrees the query latency does not grow linearly as the amount of \emph{random} I/O is bounded by the number of edge partitions. 

\mypar{Pointer-array indexing: } Recall that the {\tt pointer-array} stores location of the first out-edge of each vertex in an edge partition. To speed up access to this index, we proposed in Sec. \ref{sec:outedges} to compute an in-memory sparse index or compress the file using the Elias-Gamma coding  \cite{elias1975universal}, so that it can be loaded into memory. We evaluated these optimizations by executing a sequence of in- and out-edge queries on the {\it twitter-2010} graph. Figure \ref{fig:indexstudy} shows the mean request time: both the sparse index and the Elias-Gamma coding speed up out-edge queries significantly: by 13x and 26x respectively. But for the in-edge queries sparse indexing has almost no effect, because many in-edge queries need to touch large portions of the {\tt pointer-array} anyway. The Elias-Gamma technique speeds up the in-edge queries by almost 2xm as  now {\tt pointer-array} accesses avoid the disk.  The memory footprint of the compressed indices is 424 MB -- which easily fits in memory on our test machines -- compared to 3,383 MB for the uncompressed {\tt pointer-array} files. 

\mypar{Friends-of-Friends (FoF): } The specification of directed FoF query is as follows: qiven query vertex $u$, find vertices $W_u \subset V $ such that $\forall w \in W_u,  \, \exists v \in V  \mbox{ s. t.} \, (u, v) \in E \mbox{ and } (v, w) \in E$.  In GraphChi-DB this is implemented by first querying the out-edges of $u$ and then issuing  out-edge queries for the neighbors of $u$ simultaneously (since the edges are sorted in the edge partitions, it is more efficient to query  out-edges for several vertices simultaneously).  We compared the performance of GraphChi-DB to Neo4j using their Java API. We validated that our query implementation for Neo4j was optimal with the developers. Table \ref{table:fof} summarizes the latencies for two graphs: the small {\it livejournal} graph that can fit in memory and the much larger {\it twitter-2010} graph. For both databases we issued the same random sequence of queries, but limited the size of first-level ``friends" to 200 to avoid very long-running queries. The results show that on the smaller graph Neo4j has slightly better performance. But on the large graph, GraphChi-DB outperforms Neo4j by almost two orders of magnitude. It is clear that Neo4j's has been optimized for in-memory graphs while GraphChi-DB is optimized for disk-based access. 

\mypar{Shortest path: } Finally, we experimented with directed unweighted shortest path queries (path length was limited to five to avoid traversing the whole graph)  between two randomly chosen vertices. Shortest path query is implemented as one- or two-sided breadth-first search. To complete 5,000 such random queries on the {\it livejournal} graph, which fits in memory, Neo4j took approximately 0.04 seconds per query while GraphChi-DB used  0.2  seconds per query, on average.  Runs were executed without warm-up. With warm-up, Neo4j was even faster.  However, on the large {\it twitter-2010} graph, with a very steep power-law structure ($\alpha=1.8$), GraphChi-DB took about 8.5 seconds per query, with some queries taking up to 500 seconds. On the other hand, Neo4j failed to finish the test at all (crashing with out-of-memory exception) and used hours for some queries.  These results highlights the challenges Neo4j's design faces with very large natural graphs: the three- and four-hop neighborhoods of vertices can be massive with tens of millions of vertices. GraphChi-DB can handle such graphs gracefully: if an interim traversal frontier is very large, it can execute the traversal ``bottom-up'' as explained in Sec. \ref{sec:queries}.

%
%
%
%

\subsection{Summary of the Experimental Evaluation}

Our experiments demonstrated the scalability of our design to very large graphs and challenging read-write workloads. We showed that GraphChi-DB can execute advanced graph queries more efficiently than the leading graph database Neo4j on very large graphs and is competitive also with graphs that fit in-memory, although we have not optimized the implementation for in-memory access.  We also validated that the LSM-tree model provides excellent write performance over time. Finally, we showed that GraphChi-DB can execute analytical computation with modest decrease in query and insert performance. Due to lack of space we do not evaluate the computational capabilities in more detail but instead refer reader to \cite{graphchi:osdi2012} which contains comprehensive evaluation of the PSW algorithm.

We remark that our tests were limited to graphs with natural graph structure with degree distributions following the power-law. These kinds of graphs pose special challenges as discussed in Sec. \ref{sec:powerlaw}. With graphs with more uniform structure we expect systems like DEX and Neo4j to perform considerably better than in our tests. In general, evaluating DBMSes is hard due to differing design objectives, different optimal work-loads and complex interaction of underlying of hardware and configuration parameters. Thus, for any specific workload, the results might vary significantly. 

\vspace{-2mm}

\section{Related work}

Graph databases have been studied for at least three decades, for a survey see \cite{angles2008survey}. Perhaps best examples of modern single-computer database systems are Neo4j \cite{neo4j} and DEX \cite{dex}. Compared to this work, they do not provide powerful computational capabilities and have not been designed for extremely large graphs like GraphChi-DB.  TurboGraph \cite{turbograph} has very different design than GraphChi-DB, but also supports both database and computational operations. However, the authors do not evaluate the performance of TurboGraph\footnote{Unfortunately, we were not able to obtain TurboGraph \cite{turbograph} for evaluation, but limited experiments done by other researchers in \cite{sabrin2013mmap} are available.} in the online setting and it is unclear how efficiently it can handle graphs with edge and vertex attributes. 

Graph storage has been also studied by the Semantic Web community, for storing RDF data.  Storing RDF as a graph was first proposed in \cite{bonstrom2003storing}.  Our focus has been on graphs such as social networks and the web, but we suggest GraphChi-DB could also be used as a backend for storing RDF triples. 


\vspace{-2mm}
 
\section{Conclusions and Future work}

We have presented a new data structure, Partitioned Adjacency Lists (PAL), for storing large graphs on disk.  To enable fast access of both in- and out-edges of a vertex, without duplicating data, PAL combines strengths of standard Adjacency Lists structure with linked-list storage used by Neo4j \cite{neo4j}. The partitioning of the adjacency list enables both fast inserts and efficient disk-based analytical computation using the Parallel Sliding Windows algorithm proposed in \cite{graphchi:osdi2012}. Based on PAL, we designed GraphChi-DB and demonstrated its state-of-the-art scalability and performance on very large graphs and challenging workloads.

The PAL model is specifically designed for storing graphs, and we argue that it has several advantages over generic relational or key-value stores.  Particularly, PAL requires very small indices and thus allows handling much larger graphs on just a PC than other technologies. Also, it addresses the need to access both in- and out-edges of a vertex without duplicating data. However,  we do not necessarily agree with the argument often heard from the graph database industry that RDBMS are a fundamentally deficient technology for graph storage: for example, PAL could be used as a special table storage engine of a RDBMS, and accessed via SQL. 

\mypar{Future research: } An interesting property of PAL is that it is highly adjustable: the number of partitions, sizes of edge buffers and structure of the underlying Log-Structured Merge-tree all can be adjusted to match a target workload. We propose that it would be possible to build a system that \emph{automatically} adapts to the observed workload, using techniques from Machine Learning. Another research question is to study whether instead of using OS memory mapping for buffer management, a custom buffer manager that adapts to the structure and access patterns of the graph could provide even better performance.

\subsection*{Acknowledgements}

We thank Brandon Myers, Bill Howe and Svilen Mihailov for valuable feedback. Aapo Kyrola was supported by VMWare Graduate Fellowship.

%
%
%
%


\bibliographystyle{abbrv}

\bibliography{arxiv_graphchidb}  

\end{document}